# Entire Period Transient Stability of Synchronous Generators Considering LVRT Switching of Nearby Renewable Energy Sources

Bingfang Li, *Student Member*, *IEEE*, Songhao Yang, *Senior Member*, *IEEE*, Guosong Wang, Yiwen Hu, Xu Zhang, Zhiguo Hao, *Senior Member*, *IEEE,* Dongxu Chang, Baohui Zhang, *Fellow, IEEE*

*Abstract*—In scenarios where synchronous generators (SGs) and grid-following renewable energy sources (GFLR) are co-located, existing research, which mainly focuses on the first-swing stability of SGs, often overlooks ongoing dynamic interactions between GFLRs and SGs throughout the entire rotor swing period. To address this gap, this study first reveals that the angle oscillations of SG can cause periodic grid voltage fluctuations, potentially triggering low-voltage ride-through (LVRT) control switching of GFLR repeatedly. Then, the periodic energy changes of SGs under "circular" and "rectangular" LVRT limits are analyzed. The results indicate that circular limits are detrimental to SG's first-swing stability, while rectangular limits and their slow recovery strategies can lead to SG's multi-swing instability. Conservative stability criteria are also proposed for these phenomena. Furthermore, an additional controller based on feedback linearization is introduced to enhance the entire period transient stability of SG by adjusting the post-fault GFLR output current. Finally, the efficacy of the analysis is validated through electromagnetic transient simulations and controller hardware-in-the-loop (CHIL) tests.

## I. INTRODUCTION

THE large-scale and centralized integration of wind and solar power into the grid presents challenges like low short-circuit ratios and weak inertia support [1]-[5]. The combination of GFLR with existing SG units in the system offers a promising solution, which enhances the grid's dynamic support capability and mitigates operational challenges stemming from the inherent uncertainty of GFLR output. However, the strong power coupling between GFLRs and nearby SGs raises concerns about potential impacts on rotor angle stability in SGs. This emerging issue warrants a thorough investigation to ensure the stable operation of power systems.

Numerous scholars have investigated the rotor angle stability issues in co-located GFLR and SG systems. Some use detailed time-domain simulations for transient stability analysis [6]-[8], while others prefer simplified analytical models. On the electromechanical timescale of SG rotor motion, the phase-locked loop (PLL) dynamics of GFLR's grid-side converters (GSC) can be neglected[9]-[11]. This simplification allows GFLR to be modeled as a negative impedance[12], power source[13], or controlled current source[14], enabling classical stability analysis methods such as the direct method[15],[16], the equal area criterion[17] and phase trajectory method [18],[19] to assess system stability. Within this framework, researchers have examined how GFLR penetration, connected location, fault condition, operating mode, and voltage or frequency control affect system transient angle stability [5],[11],[13]. However, these studies often overlook the control switching of GSCs, especially the impacts of LVRT control. This oversight can lead to discrepancies between analytical results and actual system behavior.

During severe grid faults, GSCs switch to LVRT mode to prevent overcurrent and provide reactive power support[20]. Ref. [21] shows that wind farms lacking conventional support may experience voltage oscillations due to repeated LVRT control. The underlying mechanism involves a vicious circle: LVRT switching causes sudden changes in wind farm output, leading to grid voltage shifts, which in turn trigger further LVRT events. In the scenario discussed in this paper, however, SGs provide strong short-circuit capacity support. This enhances the voltage security of the sending-end system but also highlights rotor angle instability issues. Analyzing the dynamic interaction between rotor angle swings of SGs and the transient control of GFLRs is crucial to addressing this challenge.

Current research has explored the impact of GFLR's LVRT dynamics on the first-swing rotor angle stability of synchronous machines. First-swing stability refers to the system's ability to maintain stable during the first swing cycle after fault clearance[22]. This primarily depends on whether the transient energy accumulated during the fault can be effectively absorbed and dissipated after fault clearance. Ref. [23] and [24] identify two forms of rotor angle instability in GFLR stations with synchronous condensers, considering various LVRT depths and fault types. Ref. [25] studies a similar scenario in a system where GFLR is paralleled with virtual synchronous generators (VSGs). It investigates the impact of the current-limiting angle relationship between GFLR and VSG during faults on system stability, and proposes an adaptive current-limiting control method. However, these studies primarily focus on LVRT's impact during faults, overlooking the sustained influence of GFLRs on voltage source equipment after fault clearance. Ref. [26] reveals that increases in VSG rotor angle post-fault narrow the stability region of the PLL, thus adversely affecting system stability during the first swing. While this study touched upon post-fault interactions, its research perspective remained

This work was supported by The Key Science and Technology Project of CSG (GZKJXM20222178, GZKJXM20222211, GZKJXM20222213).
B. Li, S. Yang, Z. Hao, Y. Hu, X. Zhang, and B. Zhang are with Xi'an Jiaotong University, Xi'an, China (e-mail: {songhaoyang, zhghao, bhzhang}@ xjtu.edu.cn, {libingfang, huyiwen}@stu. xjtu.edu.cn). G. Wang and D. Chang are with China Southern Power Gridcsg Electric Power Research Institute.

confined to first-swing stability in the traditional sense.

Considering GFLRs' ongoing effects on subsequent swing periods after fault clearance, the issue of multi-swing stability in GFLR and SG co-located systems is gaining increasing attention. Multi-swing stability refers to the system's stability during subsequent oscillation cycles, assuming first-swing stability has been achieved. This instability often stems from new coupling effects and disturbance injections [27]. Ref. [28] graphically illustrates the acceleration and deceleration energy changes of SGs across multiple swings, revealing odd-numbered instability trends due to slow GFLR power recovery. Complementing this work, Ref. [29] provides a mathematical proof of this phenomenon. However, it treats GFLR output as an independent variable, failing to accurately account for the dynamic interactions between GFLR and main grids.

In a nutshell, existing research either solely focuses on first-swing stability or fails to account for the post-fault interactions between GFLR and the grid. Focusing on the interaction between SG rotor dynamics and GFLR output after fault clearance, this paper makes four main contributions:

1) The post-fault interaction mechanism between SG rotor angle and GFLR's LVRT control is revealed. It is found that rotor angle swings of SGs can cause voltage oscillations, which may lead to the periodic activation and deactivation of LVRT. This process, in turn, affects the rotor dynamics of SGs.

2) The entire period transient instability risks of SGs with different LVRT limit modes of GFLR are analyzed. The "circular limit" makes SGs more prone to first-swing instability, while the "rectangular limit" may cause multi-swing instability.

3) Stability boundaries in the form of critical energy are proposed for both the first-swing and multi-swing stability of SGs affected by the LVRT control switching of the GFLR.

4) Based on feedback linearization, this paper proposes a stabilization control strategy applied after fault clearance to enhance the entire period transient stability of SGs.

## II. System Modeling

### A. System Overview

Fig. 1 depicts a simplified topology of a parallel transmission system comprising GFLR and SG. They are aggregated at the point of common coupling (PCC) and connected to the receiving-end grid via AC transmission lines. The grid integration characteristics of GFLRs are primarily reflected in the interaction between the GSC and the power grid. In Fig. 1, $E_s\angle\delta$, $U_w\angle\theta$, $U_{pcc}\angle\varphi_p$, $U_g\angle 0°$ denote the voltage phasors of the sending-end SG, the converter, and the receiving-end infinite bus, respectively. $I_s\angle\varphi_s$, $I_w\angle\varphi_w$, and $I_g\angle\varphi_g$ represent the corresponding branch currents, while $Y_s$, $Y_w$, $Y_g$ are the respective branch admittances.

In Fig. 2, the $d_s$-$q_s$, $d$-$q$, and $x$-$y$ reference frames rotate counterclockwise with angular speeds $\omega_s$, $\omega_p$, and $\omega_g$. The $d_s$-axis, aligned with the SG rotor, leads the $x$-axis by an angle $\delta$, which is the SG's rotor angle. The $d$-axis of the GSC frame leads the $x$-axis by an angle $\theta$, as provided by the PLL. $\eta$ is defined as the angle by which $I_w\angle\varphi_w$ lags the $d$-axis.

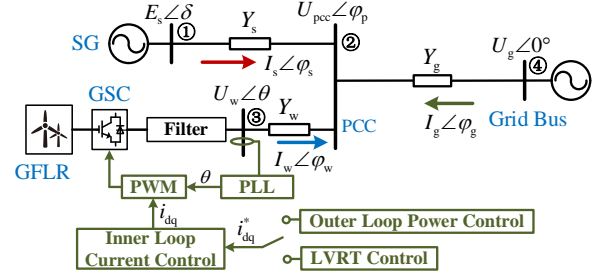

Fig. 1. Topology of the GFLR and SG co-located system.

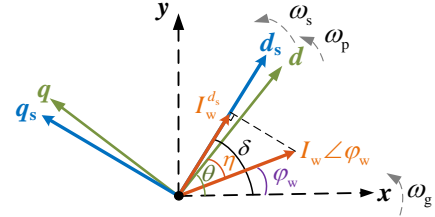

Fig. 2. Relationships of different frames in the vector space.

### B. Transient Stability Analysis Model

The GSC employs a classic cascaded control structure[30], as illustrated in Fig. 1. The control dynamics of the power electronic interface can be effectively decoupled from the rotor motion of SGs. Typically, the inner current loop (100Hz and above) and PLL (10-70Hz) operate at control bandwidths much higher than the SG rotor dynamics (approximately 1Hz), while the outer power control loop operates at an intermediate bandwidth (around 10Hz)[9]. Therefore, on the time scale of SG rotor dynamics, which is the focus of this paper, it can be assumed that the GSC control loops have reached a quasi-steady state. This means that the outer loop control of the GSC has completed its response to grid voltage changes and provided current reference values ($i_{dq}^*$). Also, the actual GSC output currents ($i_{dq}$) can be considered equal to $i_{dq}^*$. Consequently, the GSC primarily acts as a controlled current source, as determined by the outer loop or LVRT control.

Under normal grid conditions, the current reference $i_{dq}^*$ is provided by the outer power control loop. If a fault occurs causing $U_w$ to drop below the LVRT activation threshold, $U_{in}$, the LVRT is triggered. In this mode, $i_{dq}=i_{dq}^*$ is determined by the LVRT control strategy to meet the requirement for rapid current regulation. Typically, the reactive current increment of GSC should respond to voltage changes at the connection point:

$$i_q = K_q\left(0.9 - U_w\right)I_N, \qquad (1)$$

where $K_q$ is the gain coefficient of reactive current support and $I_N$ represents the rated current. Applying the reactive power priority principle, two strategies can be employed to limit GFLR active current while ensuring priority for reactive current output as per (1). The first, known as "circular limiting" and described by (2), maximizes active current while keeping the total current magnitude below $I_{max}$, the maximum allowable current of the GSC[24],[31]. This method minimizes active power losses during LVRT. The second strategy, termed "rectangular limiting" and represented by (3), directly caps the GSC active current at $i_{d,lim}^{ref}$ upon LVRT activation[20]. This method offers rapid response and simpler implementation.

$$i_d = \min\left\{\frac{P_{\text{ref}}}{U_w}, \sqrt{I_{\max}^2 - i_q^2}\right\} \quad (2)$$

$$i_d = i_{d,\lim}^{\text{ref}} \quad (3)$$

(note: $P_{\text{ref}}$ is the reference power for the outer loop control.)

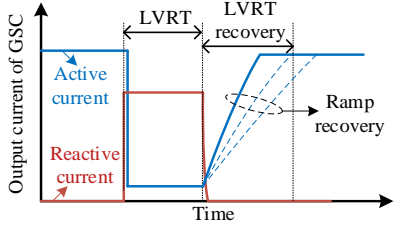

**Fig. 3.** GSC's output current under the LVRT and recovery control.

Upon fault clearance, if $U_w$ surpasses the LVRT deactivation threshold, $U_{\text{out}}$, the GSC transitions from "support mode" (described in (1)-(3)) to "recovery mode", marking the beginning of the LVRT recovery stage. To address the challenges of minimizing sudden active power changes, a common recovery strategy gradually ramps up the active current to its normal value, as depicted in Fig. 3. This approach can be expressed as:

$$i_d^* = v_d t \quad (4)$$

where $v_d$ is GFLR's active current ramp rate.

The active power of the SG can be expressed as [29]:

$$P_E = E_s U_g |Y_{sg}| \sin\delta - \alpha E_s I_w^{d_s}, \quad (5)$$

where $I_w^{d_s} = I_w \cos(\delta - \varphi_w)$ is the projection of $I_w \angle \varphi_w$ onto the $d_s$-axis;

$$|Y_{sg}| = \frac{Y_s Y_g}{Y_s + Y_g}, \quad \alpha = \frac{Y_s}{Y_s + Y_g} \in (0,1). \quad (6)$$

Neglecting the resistance component of $Y_s$ and $Y_g$, $\alpha$ is a real number. Define $P_{es} = E_s U_g / Y_{sg} / \sin\delta$ as the electromagnetic power of the SG without the influence of GFLR. $P_w = \alpha E_s I_w^{d_s}$ represents the power coupling term between the GFLR and the SG. $P_m$ is the mechanical power of the SG. Thus, the rotor dynamic equation of the sending-end SG can be expressed as:

$$\begin{cases} d\delta/dt = \omega_s - \omega_g = \omega_g \Delta\omega \\ T_J d\Delta\omega/dt = P_m - P_{es} + P_w - D\Delta\omega \end{cases} \quad (7)$$

where $P_m$, $T_J$, and $D$ denote the mechanism power, inertia time constant, and the damping coefficient of the SG, respectively.

## III. DYNAMIC INTERACTION BETWEEN SG ROTOR SWING AND POST-FAULT OUTPUT OF GFLR

After the clearance of a large disturbance, SG angle oscillations induce system voltage fluctuations, leading to changes in the GFLR current. These changes, in turn, affect SG rotor dynamics via the power coupling term, creating a closed-loop interaction: rotor angle oscillation→ voltage fluctuation→ GFLR LVRT→ GFLR output change→ rotor angle oscillation. The following analysis examines the mechanism of such an interactive process.

### A. Dynamic Process Analysis of GFLR LVRT Switching Induced by Rotor Angle Swings

The voltage at the GFLR grid connection point, denoted as $U_w \angle \theta$ in Fig. 1, can be expressed as:

$$U_w \angle \theta = \underbrace{(1-\alpha)U_g \angle 0° + \alpha E_s \angle \delta}_{U_{w1} \angle \gamma_1} + \underbrace{\beta I_w \angle (\varphi_w + 90°)}_{U_{w2} \angle \gamma_2}, \quad (8)$$

where $\beta = 1/(Y_s + Y_g) + 1/Y_w$. Fig. 4(a) illustrates the voltage phasor diagram corresponding to (8), where the vector $U_{w1} \angle \gamma_1$ starts at point $o$ and ends on the trajectory formed by the sum of vectors $(1-\alpha)U_g \angle 0°$ and $\alpha E_s \angle \delta$. Therefore, for the vector triangle $\triangle oab$, the cosine law gives:

$$U_{w1} = \sqrt{(\alpha E_s)^2 + (1-\alpha)^2 U_g^2 + 2\alpha(1-\alpha) E_s U_g \cos\delta}. \quad (9)$$

Based on Fig. 4 and (8), we know that $U_{w1} \angle \gamma_1$, $U_{w2} \angle \gamma_2$, and $U_w \angle \theta$ form the vector triangle $\triangle oac$, with $\angle oca = 90° - \eta$. Therefore, $\angle oda$ is a right angle. Using the Pythagorean theorem, $U_w$ can be expressed as:

$$U_w = U_{w2} \sin\eta + \sqrt{(\alpha E_s)^2 + (1-\alpha)^2 U_g^2 + 2\alpha(1-\alpha) E_s U_g \cos\delta - (U_{w2} \cos\eta)^2}. \quad (10)$$

As indicated by (10), the grid connection point voltage $U_w$ of the GFLR is dynamically coupled with the rotor angle $\delta$ of the SG and the current of GFLR. Within the interval $[0,\pi]$, $\delta$ varies inversely with $U_w$ in a monotonic trend. That is to say, $U_w$ decreases as $\delta$ increases. Letting $U_w = U_{\text{in}}$ (LVRT activation threshold) and $U_w = U_{\text{out}}$ (LVRT deactivation threshold), the corresponding $\delta = \delta_{\text{in}}$ and $\delta = \delta_{\text{out}}$ can be solved as follows:

$$\delta_{\text{in}} = \arccos\frac{U_{\text{in}}^2 + (\beta I_w)^2 - (\alpha E_s)^2 - (1-\alpha)^2 U_g^2}{2\alpha\beta(1-\alpha)E_s}, \quad (11)$$

$$\delta_{\text{out}} = \arccos\frac{(U_{\text{out}} - \beta I_w \sin\eta)^2 + (\beta I_{\text{out}} \cos\eta)^2 - (\alpha E_s)^2 - (1-\alpha)^2 U_g^2}{2\alpha\beta(1-\alpha)E_s}. \quad (12)$$

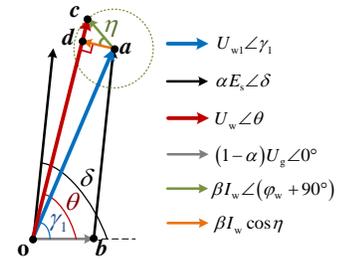

**Fig. 4.** Voltage phasor of the co-located system.

Let $\delta_m$ denote the maximum rotor angle of the SG within a given swing cycle, as shown in Fig. 5. If $\delta_{\text{in}} < \delta_m$, the GFLR will trigger its LVRT control during Phase I. Subsequently, $U_w$ rises abruptly due to the increases in $i_q^*$ and decrease in $i_d^*$. Typically, $U_{\text{out}} > U_{\text{in}}$. There will be two typical patterns after the rise of $U_w$.

**Pattern 1:** If $U_{\text{out}} < U_w$, the GFLR will immediately exit LVRT mode. However, since $\Delta\omega > 0$ at this moment, $\delta$ continues to increase according to (7), potentially causing $U_w$ to drop below $U_{\text{in}}$ again. This could lead to repeated toggling between LVRT activation and deactivation.

**Pattern 2:** If $U_{\text{in}} < U_w < U_{\text{out}}$, the converter will remain in LVRT control. Then, $\delta$ increases continuously along the orange trajectory in Fig. 5. Assuming that the SG rotor angle does not become unstable during Phase I, it enters Phase II where $\delta$ begins to decrease along the blue trajectory and $U_w$ starts to increase until $U_w > U_{out}$, at which point LVRT exits. In this scenario, $\delta_{in}$ and $\delta_{out}$ divide one swing cycle of the SG into two subsystems: the LVRT subsystem and the LVRT recovery subsystem, as shown in Fig. 7.

This analysis reveals that during one oscillation cycle after fault clearance, the converter may exhibit two switching patterns depending on the LVRT threshold settings. **Pattern 1** involves multiple rapid LVRT switches while the rotor angle remains relatively constant, where LVRT switching and voltage oscillations mutually induce each other, as described in [21]. If the number of short-term LVRT events exceeds the maximum allowable value, GFLR may disconnect from the grid. **Pattern 2** maintains the LVRT state after switching until swinging back to $\delta_{out}$ in phase II. In this scenario, **the LVRT switching frequency matches the rotor angle swing cycle of the SG**. This paper focuses on transient stability under continuous dynamic interaction between the GFLR and nearby SGs, namely the latter pattern.

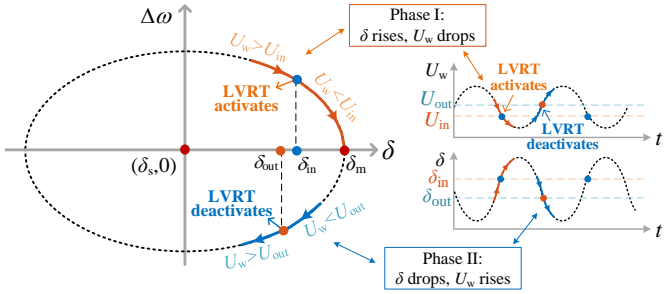

**Fig. 5.** Diagram of the correspondence between SG rotor angle swing period and LVRT switching.

*B. Impacts of GFLR Output Variations on SG Rotor Dynamics*

The previous analysis explains a single instance of LVRT activation and deactivation during the forward and return swing of the SG's rotor angle. However, whether this phenomenon will periodically recur, and thus the underlying mechanism of GFLR's impact on SG stability, still requires further analysis. To this end, we will first preliminarily analyze the impact of GFLR output variations (as disturbances) on the SG's transient energy. Following this, a systematic analysis of the SG's rotor angle stability throughout a complete oscillation cycle will be provided in the next section.

Define $P_{w*}$ as the value of $P_w$ corresponding to normal operation, which can be considered constant during the transient period (typically a few seconds). Let $\Delta P_w = P_w - P_{w*}$, system (7) is then transformed into:

$$\frac{d^2\delta}{dt^2} = \omega_g \frac{d\Delta\omega}{dt} = \frac{\omega_g}{T_J}\left[f(\delta,\Delta\omega) + \Delta P_w(t,\delta,\Delta\omega)\right], \quad (13)$$

$$f(\delta,\Delta\omega) = P_m - P_{es} + P_{w*} - D\Delta\omega. \quad (14)$$

It is known from (13) that $\Delta P_w$ can be regarded as a disturbance to the nominal system (15):

$$\frac{d^2\delta}{dt^2} = \omega_g \frac{d\Delta\omega}{dt} = \frac{\omega_g}{T_J}f(\delta,\Delta\omega). \quad (15)$$

Define $\delta_s$ as

$$\delta_s = \arcsin\frac{P_m + P_{w*}}{E_s U_g |Y_{sg}|}. \quad (16)$$

Since $f(\delta_s,0)=0$ and $\Delta P_w(t,\delta_s,0)=0$, $(\delta_s,0)$ is an equilibrium point of both the nominal system (15) and the disturbed system (13). Small disturbance analysis shows that this point is a stable equilibrium point (SEP). The classical energy function for the nominal system (15) can be chosen as:

$$V(\delta,\Delta\omega) = \frac{1}{2}T_J \Delta\omega^2 + \frac{1}{\omega_g}\int_{\delta_s}^{\delta}(P_{es} - P_{w*} - P_m)d\delta. \quad (17)$$

Eq.(17) has a clear physical significance: the first term represents the system's kinetic energy, while the second represents the potential energy. To investigate how the disturbance $\Delta P_w$ affects the system's stability, the derivative of $V(\delta,\Delta\omega)$ along the trajectory of the system (13) is given by:

$$\frac{dV}{dt} = \frac{\partial V}{\partial \delta}\frac{d\delta}{dt} + \frac{\partial V}{\partial \Delta\omega}\frac{d\Delta\omega}{dt} = -D\Delta\omega^2 + \Delta\omega\Delta P_w. \quad (18)$$

From (18), it indicates that the inherent damping always causes the system energy to dissipate due to $-D\Delta\omega^2 \leq 0$. However, the damping effect from $\Delta P_w$ is uncertain. As shown in Fig. 6, if $\Delta P_w<0$, GFLR exhibits a positive damping effect on SG's rotor in the half-plane where $\Delta\omega>0$ (Phase I and IV) of the phase plane. Conversely, a negative damping effect of GFLR is present in the half-plane where $\Delta\omega<0$ (Phase II and III). If $\Delta P_w>0$, the effects are reversed. Thus, changes in GFLR output have varying damping effects during different phases of the SG's angle oscillations, necessitating further analysis of their impact on angle stability.

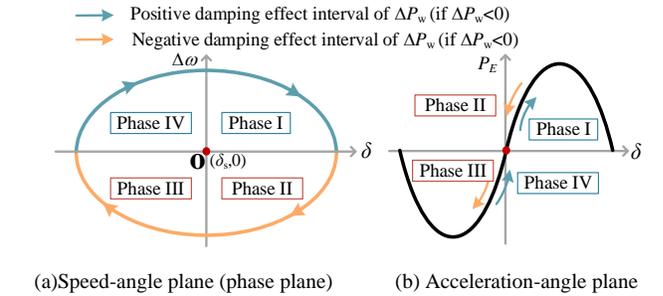

(a) Speed-angle plane (phase plane)  (b) Acceleration-angle plane
**Fig. 6.** Positive and negative damping effects of GFLR on SG's dynamics.

## IV. INSTABILITY MECHANISM AND STABILITY BOUNDARY

The previous section suggests that SG rotor angle swings might trigger repeated LVRT in GFLRs after fault clearance. The analysis in the following will focus on scenarios where SG rotor angle stability is the primary concern (Appendix A discusses the necessary conditions for this phenomenon). The instability mechanism of SGs considering their interaction is analyzed, and the stability assessment criteria are investigated.

*A. SG Stability Considering Repeated LVRT Switching*

After fault clearance, the SG's rotor angle increases. If $U_w$ drops below the LVRT threshold $U_{in}$ during this phase, the GSC will re-trigger LVRT control. This transition point, labeled as $E$ in Fig. 7, occurs when $\delta \geq \delta_{in}$. As the SG rotor angle retreats from its peak ($\delta=\delta_M$), $U_w$ climbs with the declining $\delta$. The GSC deactivates LVRT mode at point $F$ when $U_w \geq U_{out}$ and $\delta \leq \delta_{out}$.

These two critical points, $E$ and $F$, effectively split the SG's cyclic trajectory (denoted as $\Gamma$) into two segments: $\Gamma_a$, and $\Gamma_b$, representing the trajectory during LVRT and post-LVRT.

Starting from point $E$, the transient energy increment $\Delta V$ over one complete swing cycle can be expressed as:

$$\Delta V = \int_0^T \dot{V} dt = \underbrace{\frac{1}{\omega_g}\int_{EME'}\Delta P_w d\delta}_{\Delta V_{wa}} + \underbrace{\frac{1}{\omega_g}\int_{FNF'}\Delta P_w d\delta}_{\Delta V_{wb}} + \underbrace{\frac{1}{\omega_g}\left(\int_{E'F}\Delta P_w d\delta + \int_{F'E}\Delta P_w d\delta\right)}_{\Delta V_{wab}} - \underbrace{\oint_\Gamma D\Delta\omega^2 dt}_{\Delta V_D}, \quad (19)$$

where $\Delta V_{wa}$ and $\Delta V_{wb}$ represent energy increments along $\delta$-axis symmetric segments within $\Gamma_a$ and $\Gamma_b$ respectively. If two points symmetrical about the $\delta$-axis belong to $\Gamma_a$ and $\Gamma_b$ respectively, the energy increment along this path is denoted as $\Delta V_{wab}$.

Next, we will discuss the periodic energy changes of the SG under the two current limiting strategies in Section II.B.

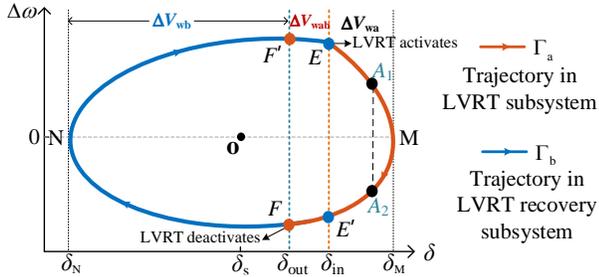
Fig. 7. Phase trajectories of SG with repeated LVRT triggering.

**1) SG stability under GSC's circular LVRT limiting**

During LVRT along the $E \to M \to E'$, changes in $\delta$ influence $U_w$. This causes the GSC's output current to adjust in response to the voltage change, which then affects $\Delta P_w$. In the scope of this study, the influence of GFLR current on $U_w$ is limited due to a strong short circuit ratio. Therefore, the magnitude of $U_w$ mainly depends on $U_{w1}$. At this point, (1) and (2) change to:

$$i_q = K_q(0.9 - U_{w1}) \quad (20)$$

$$i_d = \min\left\{P_{ref}/U_{w1}, \sqrt{I_{max}^2 - K_q^2(0.9 - U_{w1})^2}\right\} \quad (21)$$

According to (9), $U_{w1}$ changes monotonically with $\delta$ in the range of $(0,\pi)$. Therefore, it can be approximated that at two points symmetric about the $\delta$-axis on the phase plane (such as points $A_1$ and $A_2$ in Fig. 7), the magnitude and angle of the GSC output current are nearly equal. Consequently, along the trajectory $E \to M \to E'$, $\Delta P_w$ is approximately equal for the same $\delta$. Since the damping effect of $\Delta P_w$ reverses when $\Delta\omega$ changes from positive to negative, the energy changes induced by $\Delta P_w$ cancel out, resulting in $\Delta V_{wa} \approx 0$.

From point $M$ to $F$ (the LVRT deactivation point) in Fig. 7, the "circular limiting" strategy gradually reduces GSC's reactive current support to zero while smoothly transitioning active current to normal. This eliminates the need for an LVRT recovery state as shown in Fig. 4(b). Therefore, along the trajectory $F \to N \to F'$, the GSC activates the outer loop control mode, where $i_q=0$ and $i_d=P_{ref}/U_w$. Similarly, the energy increments caused by $\Delta P_w$ along $F \to N$ and $N \to F'$ cancel each other out, so $\Delta V_{wb} \approx 0$.

Despite different GSC control modes in $E' \to F$ and $F' \to E$ segments, the slight variations in $U_w$ near LVRT activation and deactivation points lead to small changes in GSC current. This results in similar $\Delta P_w$ values in $E' \to F$ and $F' \to E$, making $\Delta V_{wab}$ negligible. Thus, LVRT reactivation minimally impacts SG's periodic energy under the "circular limiting".

Nevertheless, if the reactive current coefficient $K_q$ (in (1)) is small, the drop in $U_w$ during the first swing prompts the GSC to increase active current output up to $I_{max}$ to maintain steady power as much as possible. This results in a positive $\Delta P_w$, which, combined with $\Delta\omega>0$ at this stage, introduces negative damping from the GFLR. Thus, the SG's first swing stability is exacerbated under these conditions.

**2) SG stability under GSC's rectangular LVRT limiting**

The "rectangular limiting" strategy imposes tighter constraints on active current, leading to a lower $I_w^{ds}$ during LVRT. Consequently, this approach yields a significantly smaller $\Delta P_w$ than the "circular limiting". The reduced $\Delta P_w$ enhances the positive damping effects along the trajectory $E \to M$, which is more conducive to the SG's first-swing stability.

To analyze periodic energy variations, we define $\Delta P_{wa}$ and $\Delta P_{wb}$ as the $\Delta P_w$ values along trajectories $\Gamma_a$ and $\Gamma_b$ respectively. First, we analyze the magnitude of $\Delta V_{wa}$. Similar to "circular limiting", for two points symmetrical about the $\delta$-axis, $\Delta P_{wa} \approx \Delta P_{wb}$, leading to $\Delta V_{wa} \approx 0$. Consequently, the magnitude of $\Delta V_w$ is primarily determined by $\Delta V_{wab}$ and $\Delta V_{wb}$:

$$\Delta V_w = \underbrace{\frac{1}{\omega_g}\int_{\delta_{in}}^{\delta_{out}}(\Delta P_{wa} - \Delta P_{wb})d\delta}_{\Delta V_{wab}} + \underbrace{\frac{1}{\omega_g}\left(\int_{\delta_{out}}^{\delta_N}\Delta P_{wb}d\delta + \int_{\delta_N}^{\delta_{out}}\Delta P_{wb}d\delta\right)}_{\Delta V_{wb}} \quad (22)$$

Next, we analyze $\Delta V_{wab}$. Under the constraints of the current limiting, $\Delta P_{wa}$ on $E' \to F$ is smaller than $\Delta P_{wb}$ on $F' \to E$. It is deduced from (22) that if $\delta_{out} < \delta_{in}$, then $\Delta V_{wab} > 0$; conversely, if $\delta_{out} > \delta_{in}$, then $\Delta V_{wab} < 0$. Therefore, $\Delta V_{wab}$ primarily depends on LVRT switching conditions specified in (11) and (12).

For the $F \to N \to F'$ interval, if the slow current recovery strategy is adopted, the GSC's current gradually returns to normal levels, making $\Delta P_w$ time-dependent. During recovery, $\Delta P_w(t)$ remains negative but increases over time, behaving monotonically with respect to $\delta$ in both $F \to N$ and $N \to F'$ intervals. According to the Mean Value Theorem for Integrals, there exist constants $C_1$ and $C_2$, with $C_1 < C_2 \leq 0$, that satisfy:

$$\int_{\delta_{out}}^{\delta_N}\Delta P_{wb}(t)d\delta = C_1(\delta_N - \delta_{out}). \quad (23)$$

$$\int_{\delta_N}^{\delta_{out}}\Delta P_{wb}(t)d\delta = C_2(\delta_{out} - \delta_N). \quad (24)$$

Adding (23) and (24) gives the system energy increment $\Delta V_{wb}$ along the trajectory $F \to N \to F'$:

$$\Delta V_{wb} = (C_2 - C_1)(\delta_{out} - \delta_N)/\omega_g > 0. \quad (25)$$

Therefore, the slow recovery strategy results in an accumulation of transient energy. Even if $\Delta V_{wab} < 0$, $\Delta V_w$ may still be positive due to $\Delta V_{wb} > 0$ under the slow recovery. If $\Delta V_w - \Delta V_D > 0$, cyclic energy accumulation occurs. This, in turn, exacerbates the magnitude of its rotor angle swings. A more pronounced angle swing naturally leads to more severe voltage oscillation, thereby triggering LVRT switching in the subsequent cycle and leading to further accumulation of transient energy. Once this vicious cycle in Fig. 8 is established, the GFLR's LVRT is periodically triggered, and additional energy accumulates in each cycle, resulting in multi-swing angle instability in the SG.

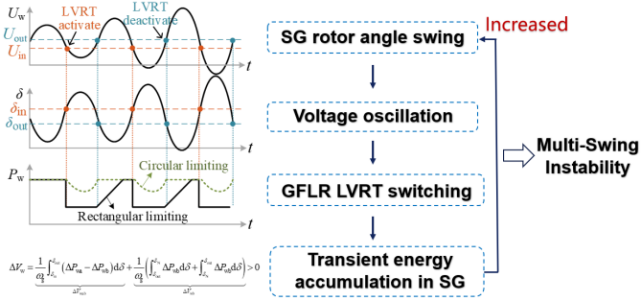

**Fig. 8.** The vicious cycle driving SG multi-swing instability.

**In a nutshell, if the GFLR uses "circular limiting" with a low reactive support coefficient, the SG has a higher risk of first-swing instability. Conversely, if the GFLR uses "rectangular limiting" with a slow recovery strategy, there is a risk of multi-swing instability in the nearby SG.** For a multi-machine system, the sending-end SGs can be coherently aggregated. The injected current $I_w\angle\varphi_w$ at the PCC in Fig. 1 can be considered as the combined output of multiple GFL converters. Thus, the conclusions of this analysis also apply.

*B. Stability Boundaries in the Form of Critical Energy*

The following discussion explores the first-swing and multi-swing stability boundaries for SGs in the form of critical energy, considering the control dynamics of the nearby GFLR cluster.

**1) First-Swing Stability Boundary**

Based on Section III.B, if $P_w$ reaches its maximum value ($P_w=\alpha E_s I_{max}$), the GFLR exerts the greatest negative damping effect during the first-swing phase (Phase I) after fault clearance. This results in the conservative first-swing stability boundary. Therefore, the first-swing critical energy of the SG, denoted as $V_f^{max}$, is as follows:

$$V_f^{max} = \frac{1}{\omega_g}\int_{\delta_s}^{\pi-\delta_s}\left(E_s U_g |Y_{sg}|\sin\delta - \alpha E_s I_{max} - P_m\right)d\delta \quad (26)$$

The SG maintains first-swing stability if the energy at fault clearance, denoted as $V_c$, is less than $V_f^{max}$.

**2) Multi-Swing Stability Boundary**

Sudden output variations at the moment of GFLR LVRT control switching, as well as time-varying GFLR currents during the slow recovery phase, may result in periodic energy accumulation in SG. This potentially triggers multi-swing instability. The actual multi-swing stability boundary forms an unstable limit circle ($\Gamma_{u1}$). The proof is provided in Appendix B. Since $\Gamma_{u1}$ is difficult to analyze theoretically, a more conservative stability boundary is needed, which is constructed based on two main principles:

1) Preventing GFLR from re-entering the LVRT switching state after fault clearance.

2) Maximizing transient energy accumulation during the LVRT recovery period.

To prevent GFLR from re-entering the LVRT switching state after fault clearance, SG's rotor angle should be less than $\delta_{in}$. Thus, the critical energy of SG should be less than the energy in (26) corresponding to ($\delta_{in}$, 0), denoted as $V_{m1}^{max}$:

$$V_{m1}^{max} = \frac{1}{\omega_g}\int_{\delta_s}^{\delta_e}\left(E_s U_g |Y_{sg}|\sin\delta - \alpha E_s I_{max} - P_m\right)d\delta \quad (27)$$

where

$$\delta_e = \begin{cases} \delta_{in}, \delta_{in} < \pi - \delta_s \\ \pi - \delta_s, \delta_{in} > \pi - \delta_s \end{cases} \quad (28)$$

Even if GFLR does not re-enter LVRT mode after fault clearance, its LVRT recovery control will still introduce time-varying disturbances to SG. To account for this, the energy increment during the LVRT recovery period, $\Delta V_{wr}$, should be calculated along the path of maximum transient energy for SG. Despite the complexity of disturbances caused by GFLR after fault clearance, $\Delta P_w$ has upper and lower limits. From (18), it is known that if $\Delta\omega<0$ (phases I and IV), taking the minimum value of $\Delta P_w$ maximizes energy accumulation; if $\Delta\omega>0$ (phases II and III), taking the maximum value of $\Delta P_w$ maximizes energy accumulation. According to (5), $0<P_w<\alpha E_s I_{max}$, so the maximum ($\Delta P_w^{max}$) and minimum ($\Delta P_w^{min}$) values of $\Delta P_w$ can be expressed as:

$$-P_{w*} = \Delta P_w^{min} < \Delta P_w < \Delta P_w^{max} = \alpha E_s I_{max} - P_{w*} \quad (29)$$

Therefore, as shown in Fig. 9, $\Delta V_{wr}$ can be maximized along path $\Gamma_b$. Let $V_m^{max} = V_{m1}^{max} - \Delta V_{wr}$. This represents the equivalent stability boundary in terms of energy and can be expressed as follows:

$$V_m^{max} = V_{m1}^{max} - \frac{1}{\omega_g}\int_{\delta_N}^{\delta_M}\left(\alpha E_s I_{max}\right)d\delta \quad (30)$$

where $\delta_M$ and $\delta_N$ can be determined by solving (31) and (32):

$$V_c = \int_{\delta_s}^{\delta_M}\left(P_{es} - \alpha E_s I_{max} - P_m\right)/\omega_g d\delta \quad (31)$$

$$\int_{\delta_Z}^{\delta_M}\left(P_{es} - P_m\right)d\delta = \int_{\delta_N}^{\delta_Z}\left(P_m - P_{es}\right)d\delta \quad (32)$$

where $\delta_Z$ is the $\delta$ value at the intersection of $P_E^{min}$ and $P_m$. If $V_c$ is less than $V_m^{max}$, SGs can be considered multi-swing stable.

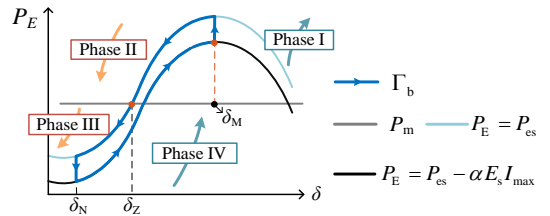

**Fig. 9.** Path of maximum energy increment during the LVRT recovery phase on the $P_E$-$\delta$ plane.

## V. CONTROLLER DESIGN

Next, we design an additional controller using feedback linearization. This controller adjusts the GFLR's output after fault clearance to transform its negative damping effect on the rotor motion of SGs into positive damping.

## A. Controller Design Based on Feedback Linearization

After applying an additional current controller, $I_{add}$, in the post-fault period, the rotor motion equation of the SG becomes:

$$\begin{cases} d\delta/dt = \omega_g \Delta\omega_* \\ T_J d\Delta\omega_*/dt = P_m - P_{E*} - D\Delta\omega_* + \alpha E_s I_{add} \end{cases}, \quad (33)$$

where $\delta$, $\Delta\omega_*$, and $P_{E*}$ represent the SG's controlled rotor angle, angular velocity deviation, and electromagnetic power, respectively. According to the feedback linearization theory, $I_{add}$ can be designed as

$$I_{add} = (P_{E*} - P_m)/(\alpha E_s) - K_1 \Delta\omega_* - K_2(\delta_* - \delta_d). \quad (34)$$

In (34), the nonlinear term $P_m - P_{E*}$ is eliminated and simultaneously introducing a linear state feedback term $-K_1\Delta\omega_* - K_2(\delta_* - \delta_d)$, where $K_1$ and $K_2$ function as the tuning parameters. Accordingly, the following linear closed-loop system can be obtained:

$$\begin{bmatrix} \dot{e} \\ \dot{\sigma} \end{bmatrix} = \begin{bmatrix} 0 & \omega_g \\ -K_2/T_J & -(K_1+D)/T_J \end{bmatrix} \begin{bmatrix} e \\ \sigma \end{bmatrix}, \quad (35)$$

where $e = \delta_* - \delta_d$ and $\sigma = \Delta\omega_*$ are the track errors.

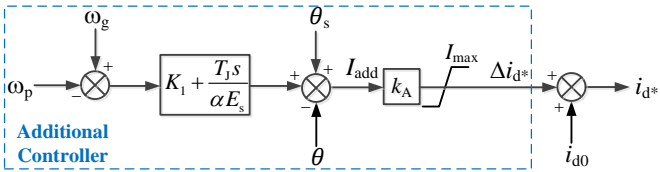

**Fig. 10.** Block diagrams of proposed control method.

## B. Stability Proof and Parameter Selection

The Lyapunov function, specifically selected for this controlled system, is established as follows:

$$V(e,\sigma) = 0.5e^2 + 0.5T_J\sigma^2. \quad (36)$$

The only equilibrium point of system (35) is (0,0). According to (36), $V(e,\sigma)$ is positive definite because it satisfies $V(e,\sigma) \geq 0$ and is equal to zero only when $V(0,0)=0$. The derivative of $V(e,\sigma)$ with respect to time is:

$$\dot{V}(e,\sigma) = \frac{\partial V}{\partial e}\dot{e} + \frac{\partial V}{\partial \sigma}\dot{\sigma}$$
$$= (1-K_2)e\sigma - (K_1+D)\sigma^2 \quad (37)$$

According to (37), $dV/dt < 0$ is valid if $K_2=1$ and $K_1>0$. Therefore, the system will achieve asymptotic stability at $(\delta_*, \Delta\omega_*) = (\delta_d, 0)$ under these conditions.

In (34), $E_s$ can be approximated as a constant under excitation control. $\alpha$ can be estimated based on the system's pre-fault parameters, which is valid because the system parameters after fault clearance do not differ significantly from those before the fault. Due to the close electrical distance between GFLR and sending-end SG, once the GSC's PLL converges quickly after fault clearance, its frequency and phase can approximate the SG's angular velocity and rotor angle. Specifically, $\Delta\omega_* \approx \omega_p - \omega_g$ and $\delta_* - \delta_d \approx \theta - \theta_s$, where $\theta_s$ is the angle of the pre-fault PLL SEP. Neglecting the inherent damping (which results in a more conservative control approach), $P_m - P_{E*}$ can be estimated as $T_J d\Delta\omega_*/dt$. After fault clearance, the GFLR's reactive current support is withdrawn, so $I_{add}$ is adjusted through the GFLR's active current. Let the command for the adjustment in active current be $\Delta i_{d*}$. Thus, the final control law is given by

$$\Delta i_d^* = k_A \left[ -\left(k_1 + \frac{T_J s}{\alpha E_s}\right)(\omega_p - \omega_g) - (\theta - \theta_s) \right], \quad (38)$$

where $k_A$ is set to be greater than 1 to amplify the control effect since $I_{add} = \Delta i_{d*}\cos(\delta - \theta)$ according to (5). Adding $\Delta i_{d*}$ to the normal current reference $i_{d0}$ gives the current reference $i_{d*}$ under the proposed control. This control block is shown in Fig. 10. The controller activates when the LVRT control quits. Control duration can be set to a typical rotor transient time, within several seconds.

It is important to note that under the proposed strategy, the GFLR's active current decreases if $\omega_p$ and $\theta$ increase, stabilizing the PLL. Additionally, once the control target is achieved, it indicates that the GFLR output has returned to normal. Thus, this additional controller not only prevents adverse effects on SG power angle stability but also supports PLL grid stability and timely recovery of active current after LVRT ends. However, due to the limited adjustment of $\Delta i_{d*}$, global stabilization of the system can be achieved only if the upper and lower limits of $\Delta i_{d*}$ are not reached during the control period.

## VI. NUMERICAL RESULTS

### A. Test System 1: Multi-Machine System Simulation

A simplified equivalent model of one actual grid was built on the PSD-BPA simulation platform. Simulation parameters are provided in Appendix C. Topology of test system 1 is shown in Fig. 11. A permanent three-phase metallic short circuit occurs on one of the three outgoing lines at $t=1$s. To verify the phenomenon of SG multi-swing instability, the fault is set near the PCC on the line for 120 ms. Test conditions for Cases 1-4 are listed in TABLE I.

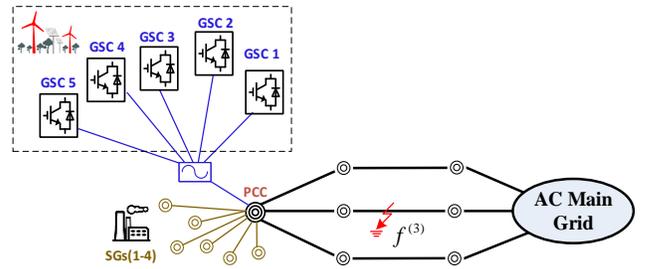

**Fig. 11.** Topology of the test system 1.

Fig. 12 presents the simulation results, showing SG rotor angles, GFLR grid-connected voltages, and GFLR output current. After fault clearance, $U_w$ shows an opposite trend to $\delta$. When $U_w$ drops below the LVRT threshold $U_{in}$ as $\delta$ increases, the GFLRs' active current decreases while the reactive current increases. Comparing Cases 1 and 2, "rectangular limiting" enhances SG first-swing stability more effectively than "circular limiting." However, in Case 3, increasing the LVRT deactivation threshold, and in Case 4, changing the recovery strategy from immediate to ramp, both increase the risk of SG multi-swing instability compared to Case 2. This indicates that repeated LVRT along with specific switching conditions and slow recovery strategies can worsen SG multi-swing stability, confirming the theoretical analysis in Section IV. A.

TABLE I. CONDITIONS AND RESULTS OF TEST SYSTEM 1

| Case | 1 | 2 | 3 | 4 |
|---|---|---|---|---|
| Limit type | circular | rectangular | rectangular | rectangular |
| reactive coefficient ($K_q$) | 2 | 2 | 2 | 2 |
| Active current limit ($i_{d,lim}^{ref}$) | GSC1:25%; GSC2:40%; GSC3:30%; GSC4:20%; GSC5:35% | | | |
| LVRT threshold ($U_{in}, U_{out}$) | 0.9,0.91 | 0.9,0.91 | 0.9,0.95 | 0.9,0.95 |
| Recovery rate ($v$) | ∞ | ∞ | ∞ | GSC1-5: 2.5, 2.5,5,1.5,2A/s |
| Result | first-swing unstable | stable | oscillation | multi-swing unstable |

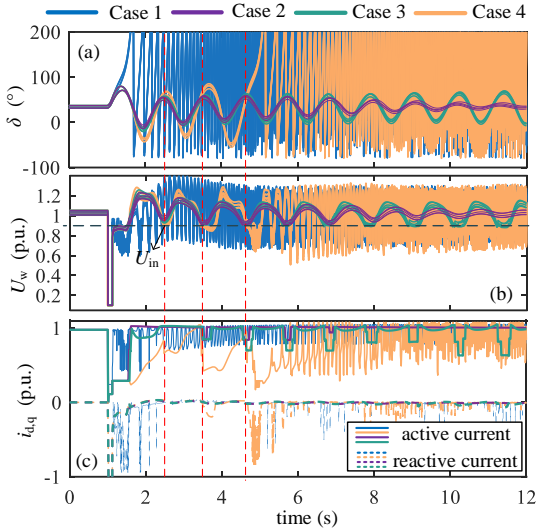

**Fig. 12.** Simulation results of Cases 1-4. (a) rotor angle curves of SGs; (b) grid connection point voltage of GFLRs; (c) sum of active current of GFLRs.

Next, we will verify the stability assessment criteria proposed in Section IV. B. Cases 5, 6, and 7 represent three scenarios: "circular limiting", "rectangular limiting" with immediate recovery, and "rectangular limiting" with ramp recovery, respectively. In these cases, the ranges for $K_q$, $U_{in}$, $U_{out}$, $i_{d,lim}^{ref}$, and $v$ are [1.5, 3], [0.8, 0.9], [0.85, 0.95], [20%, 90%], and [1 A/s, 5 A/s], respectively. The fault location is set at the midpoint of the transmission line. Fault durations $t_c$ are set to 20, 40, 60, and 100 ms to evaluate stability criteria. For each $t_c$, adjust $K_q$, $U_{in}$, $U_{out}$, $i_{d,lim}^{ref}$, and $v$ within the specified ranges to observe any scenarios. The resulting rotor angle curves for representative stable and unstable scenarios are shown in Fig. 13.

TABLE II. shows the critical energy calculations for first-swing ($V_f^{max}$) and multi-swing ($V_m^{max}$) instability. Let $V_d$ represent the transient energy of SGs at fault clearance. If $V_d < V_f^{max}$ and $V_d < V_m^{max}$, the system maintains first and multi-swing stability; otherwise, it faces the risk of instability. As shown in TABLE II., when the criteria indicate the system is stable at $t_c$ =20ms, simulation results confirm this stability. At $t_c$=60 and 100 ms,

both the criteria and simulation predict instability. However, at $t_c$=40 ms, the criteria suggest first-swing stability but multi-swing instability, while the simulation indicates stability, revealing that the criteria are conservative.

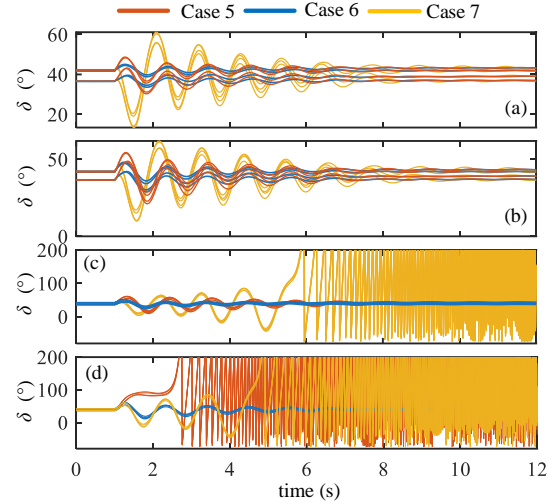

**Fig. 13.** Rotor angle curves in Cases 5-7. (a) $t_{cr}$=20ms; (b) $t_{cr}$=40ms; (c) $t_{cr}$=60ms; (d) $t_{cr}$=100ms.

TABLE II. STABILITY ASSESSMENT AND SIMULATION RESULTS

| $t_{cr}$ (ms) | Estimated Critical Energy ($V_f^{max}$, $V_m^{max}$) | | Fault Clearance Energy ($V_d$) | | Stability Assessment | Simulation result |
|---|---|---|---|---|---|---|
| | First-swing | Multi-swing | Case5 | Case 6,7 | | |
| 20 | 1.76 | 0.91 | 0.31 | 0.29 | stable | stable |
| 40 | 1.76 | 0.57 | 1.19 | 1.06 | unstable | stable |
| 60 | 1.76 | <0 | 2.69 | 2.33 | unstable | unstable |
| 100 | 1.76 | <0 | 8.97 | 3.17 | unstable | unstable |

### B. Test System 2: Controller Hardware-in-the-Loop Test

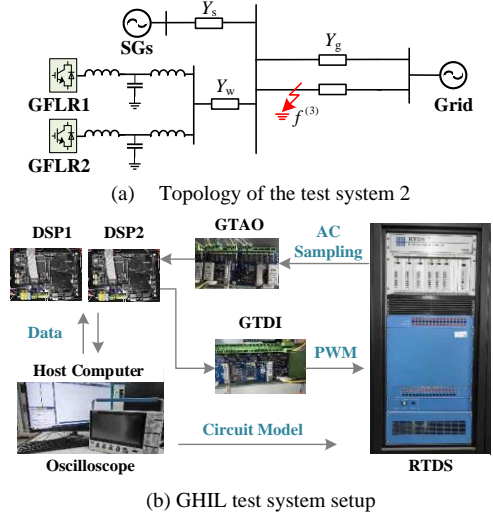

**Fig. 14.** Configuration of the CHIL platform.

To verify the theoretical analysis, a CHIL platform is constructed, as shown in Fig. 14. The main circuit shown in Fig. 14(a) is simulated in RTDS, while the controllers of the GFLRs are implemented on a DSP-TMS320F28335 board. Data exchange between the RTDS and DSPs is facilitated through Giga-Transceiver Analog Output (GTAO) and Giga-Transceiver Digital Input (GTDI) interface boards. At the sending end, 10 converters are divided into two groups (GFLR1 and GFLR2), each with identical parameters. System parameters are provided in Appendix C TABLE V. At $t$=1.7s, a three-phase short circuit occurs near the receiving end of one of the double-circuit transmission lines. The protection relay then isolates the faulty line, followed by a reclosing action to reconnect it. The proposed control activates when LVRT deactivates and is set to last for 4 seconds.

TABLE III. CONDITIONS AND RESULTS OF TEST SYSTEM 2

|  | Case 8 | | Case 9 | | Case 10 | |
| --- | --- | --- | --- | --- | --- | --- |
|  | GFLR1 | GFLR2 | GFLR1 | GFLR2 | GFLR1 | GFLR2 |
| Limit type | circular | | rectangular | | rectangular | |
| Steady power | 60MW | 47MW | 60MW | 47MW | 60MW | 47MW |
| $U_{in}, U_{out}$ | 0.87, 0.95p.u | 0.86, 0.9p.u. | 0.87, 0.95p.u | 0.9, 0.95p.u | 0.87, 0.95p.u | 0.86, 0.9p.u. |
| $K_q$ | 1.5 | 2.5 | 1.5 | 2.5 | 1.5 | 2.5 |
| $i_{d,lim}^{ref}$ | 15% | | 15% | | 15% | |
| $v$ | 2p.u./s | 1p.u./s | 2p.u./s | 1p.u./s | 2p.u./s | 1p.u./s |
| $k_1, k_A$ | 2, 1.5 | 2, 2 | 2, 1.5 | 2, 2 | 2, 1.5 | 2, 2 |
| Fault duration | 100ms | | 200ms | | 150ms | |
| Reclose | After 3s | | After 2s | | After 3s | |

Cases 8, 9, and 10 compare and verify the effectiveness of the proposed control method in suppressing the first-swing instability and multi-swing instability of the SG. The test conditions are listed in TABLE III. The results are presented in Fig. 15, and 16. Fig. 15 demonstrates that the SG experiences first-swing instability under the circular LVRT limiting control. From fault clearance to loss of synchronism, Fig. 15 (g) shows that the GFLRs exert a negative damping effect, leading to transient energy accumulation in the SG, denoted as $E_−$. However, when the proposed control method is applied, GFLRs provide a positive damping effect during the first swing, resulting in energy dissipation in the SG, denoted as $E_+$, as illustrated in Fig. 15 (h).

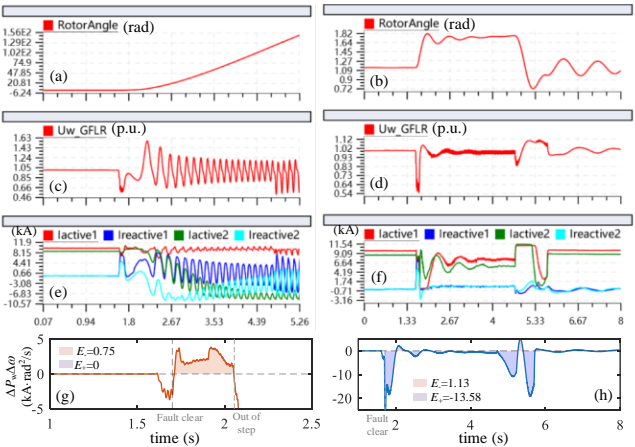

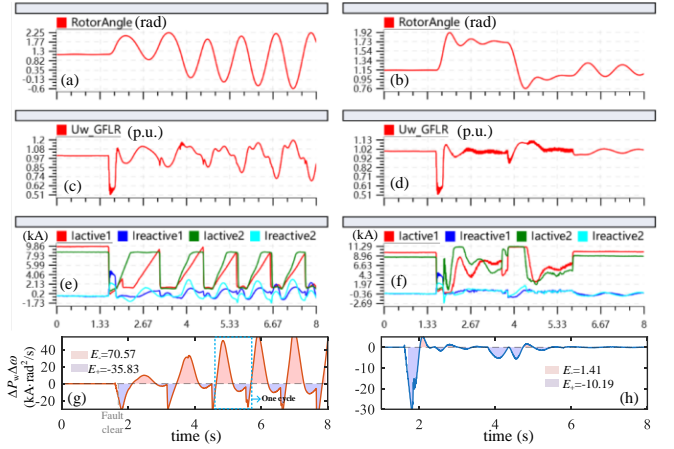

**Fig. 16.** Simulation results of Cases 9. (a) SG rotor angle ($δ$) without control; (b) SG rotor angle ($δ$) with control; (c) GFLR grid-connection voltage ($U_w$) of without control; (d) GFLR grid-connection voltage ($U_w$) with control; (e) active and reactive current of GFLR without control; (f) active and reactive current of GFLR with control; (g) GFLR energy increment without control; (h) GFLR energy increment with control.

Fig. 16 (g) and Fig. 17 (g) reveal that compared to circular limiting, rectangular LVRT limiting reduces the transient energy of the SG during the first swing after fault clearance, thus enhancing the first-swing stability. However, it may increase the risk of oscillations or instability in later swings. In Case 9 (Fig. 16), after the fault line is reclosed, the oscillation of $δ$ causes $U_w$ to oscillate inversely, periodically triggering the LVRT control of the GFLRs. Fig. 16 (g) shows that in each oscillation cycle, the negative damping effect from GFLRs leads to more energy accumulation than the positive damping dissipates, resulting in ongoing oscillations despite physical damping. In Case 10 (Fig. 17), the SG becomes unstable during the third swing due to delayed reclosing. With the proposed control, the SG rotor angle stabilizes in both Cases 9 and 10, demonstrating the effectiveness of the proposed control.

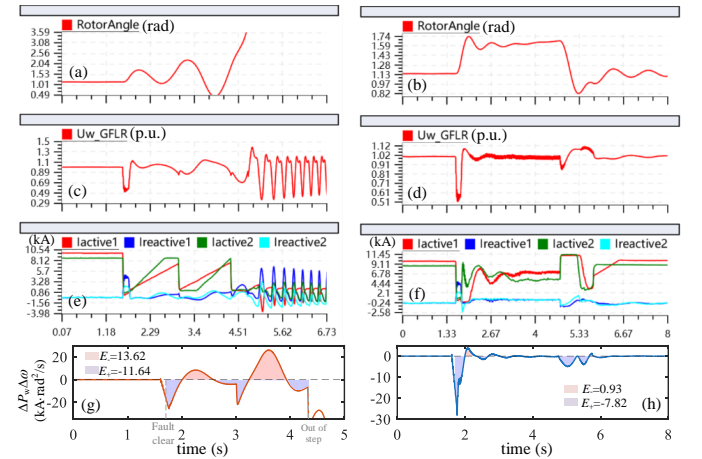

**Fig. 15.** Simulation results of Cases 8. (a) SG rotor angle ($δ$) without control; (b) SG rotor angle ($δ$) with control; (c) GFLR grid-connection voltage ($U_w$) of without control; (d) GFLR grid-connection voltage ($U_w$) with control; (e) active and reactive current of GFLR without control; (f) active and reactive current of GFLR with control; (g) GFLR energy increment without control; (h) GFLR energy increment with control.

**Fig. 17.** Simulation results of Case 10. (a) SG rotor angle ($δ$) without control; (b) SG rotor angle ($δ$) with control; (c) GFLR grid-connection voltage ($U_w$) of without control; (d) GFLR grid-connection voltage ($U_w$) with control; (e) active and reactive current of GFLR without control; (f) active and reactive current of GFLR with control; (g) GFLR energy increment without control; (h) GFLR energy increment with control.

## VII. Discussion

This section aims to deepen the understanding of this multi-swing instability problem from the dual perspectives of scenarios and solutions: first, investigating whether the observed phenomenon would persist if SG were replaced by grid forming (GFM) devices, and second, discussing more alternative strategies for enhancing the SG's stability.

### A. Similar Scenarios

As GFM technology matures, co-located systems of GFLRs and GFM devices are poised to become mainstream in the future. Given that Virtual Synchronous Generator (VSG)-type GFM converters exhibit external characteristics similar to those of SGs, they could similarly manifest the multi-swing instability behavior investigated in this paper. Nevertheless, the highly flexible and adjustable virtual damping of GFM converters offers the potential to significantly reduce their multi-swing instability risk compared to SGs.

However, due to their limited overcurrent capability, GFM converters may remain current-limiting even after the fault is cleared[33]. This condition not only compromises their own transient stability[34] but also causes a significant reduction in grid strength. As a result, the risk of PLL transient instability for the GFLRs would be greatly increased. During such periods, the dynamic interactions among GFM converters, PLLs, and GFLR's LVRT control become considerably complex. Whether the multi-swing instability needs to be considered in such scenarios deserves further in-depth investigation.

### B. Alternative Solutions

In addition to the control proposed in this paper, modifying LVRT parameters and adjusting the SG's own control strategies are also alternative solutions.

#### 1) Modifying LVRT Parameters

According to the analysis in this paper, adjusting the LVRT parameter could be considered a potential solution. However, this approach faces several practical challenges. First, LVRT strategies typically prioritize device overcurrent limits and grid code compliance instead of stability objectives. Second, simultaneously addressing both first-swing and multi-swing stability through such adjustments is challenging due to their contradictory requirements. For instance, while greater active current reduction or slower recovery during LVRT may aid first-swing stability, it can concurrently increase the risk of multi-swing instability. Furthermore, modifying LVRT thresholds also introduces additional complications. A narrow deadband between LVRT activation ($U_{in}$) and deactivation ($U_{out}$) thresholds can lead to undesirable LVRT toggling, dictated by the rapid inner current control loop. Conversely, a wider deadband may, as this study shows, trigger repeated LVRT events on the SG rotor dynamic timescale that are of concern. Consequently, relying on LVRT parameter adjustments is not recommended as a primary solution to this specific stability problem.

#### 2) SG Control Alternatives

Alternative SG control strategies include disconnecting SG if unstable or enhancing its excitation. While disconnecting the SG appears straightforward, this approach sacrifices the SG's contribution to grid strength and risks inducing secondary large disturbances in PLLs of GFLRs.

Enhanced excitation control, on the other hand, offers a more sophisticated solution. Beyond its primary role in voltage regulation, it can be strategically used to improve angle stability. Under GFLR influence, this necessitates not only improving first-swing stability but also implementing comprehensive four-quadrant regulation (Fig. 6). A promising direction integrates Lyapunov-based energy dissipation criteria with supplementary excitation, which allows for real-time active damping based on rotor speed and power imbalance. However, practically implementing full-period supplementary excitation requires coordination with overvoltage limits and protection schemes (especially transformer differential protection) to avoid undesired trips resulting from increased excitation currents.

## VIII. Conclusion

On the transient stability of the SG and GFLR co-located system, this paper extends the analysis beyond the first swing stability to encompass the entire dynamic process. It reveals a periodic forced instability phenomenon in SGs induced by a "vicious cycle" after fault clearance. Specifically, the SG's rotor angle oscillations trigger voltage fluctuations, which can re-trigger the GFLR's LVRT activation and deactivation. While LVRT limits can enhance SG's first-swing stability, they may promote transient energy accumulation in subsequent swings. Ultimately, more severe rotor angle swings trigger another round of this switching process, potentially leading to SG multi-swing instability. Under worst-case scenarios, the stability margin for this GFLR-induced multi-swing instability may become narrower than that of the first-swing stability, demanding dedicated mitigation strategies. To address this, the paper proposes an additional GFLR current controller based on feedback linearization theory, which effectively suppresses GFLR-induced multi-swing instability while simultaneously improving first-swing stability.

## Appendix

### A. System Conditions for the Studied Scenarios

$\alpha$ indicates the strength of the electrical connection between the SG and the GFLR. As SG capacity increases or grid connection impedance decreases, $Y_s$ increases, resulting in a larger $\alpha$. Next, we discuss the range of $\alpha$ in the scenarios examined in this paper.

#### 1) How $\alpha$ changes GFLR current's impact on SG rotor speed

According to (7), $P_w$ increases as $\alpha$ increases. This indicates that GFLR output changes have a more significant impact on the rotor dynamics of SG with a higher $\alpha$.

#### 2) How $\alpha$ changes the impact of $\delta$ on $U_w$

The sensitivity of $U_w$ to $\cos\delta$ (or $\delta$) can be expressed as:

$$\frac{\partial U_w}{\partial(\cos\delta)} = \frac{\alpha(1-\alpha)E_s U_g}{\sqrt{(\alpha E_s)^2 + (1-\alpha)^2 U_g^2 + 2\alpha(1-\alpha)E_s U_g \cos\delta}} = g(\alpha) \quad (39)$$

From (39), we see that $g(\alpha) \geq 0$, with $g(\alpha)=0$ only when $\alpha=0$ or $\alpha=1$. Thus, $g(\alpha)$ reaches a maximum within the interval $0<\alpha<1$. By setting $g'(\alpha)=0$ assuming $E_s \approx U_g$ in per unit values, we find that $g'(\alpha)=0$ at $\alpha=0.5$. At this point, $U_w$ is most sensitive to changes in $\delta$.

### 3) How $\alpha$ changes the impact of GFLR Current on $U_w$

After fault clearance, the GFLR current is is mainly composed of active current. The sensitivity of $U_w$ to $i_d$ can be expressed as:

$$\frac{\partial U_w}{\partial i_d} = -\frac{\beta^2 i_d}{\sqrt{(\alpha E_s)^2 + (1-\alpha)^2 U_g^2 + 2\alpha(1-\alpha)E_s U_g \cos\delta - (\beta i_d)^2}} = r(Y_s) \quad (40)$$

Differentiating equation (40) with respect to $Y_s$ yields:

$$r'(Y_s) = 2i_d \left(\frac{1}{Y_g+Y_s}+\frac{1}{Y_w}\right) \Big/ \left[(Y_g+Y_s)^2 \sqrt{\alpha^2 E_s^2 + 2\alpha(1-\alpha)E_s U_g \cos\delta + (1-\alpha)^2 U_g^2 - i_d^2\left(\frac{1}{Y_g+Y_s}+\frac{1}{Y_w}\right)^2}\right]$$
$$+ i_d^3 \left(\frac{1}{Y_g+Y_s}+\frac{1}{Y_w}\right)^3 \Big/ \left[(Y_g+Y_s)^2 \left(\alpha^2 E_s^2 + 2\alpha(1-\alpha)E_s U_g \cos\delta + (1-\alpha)^2 U_g^2 - i_d^2\left(\frac{1}{Y_g+Y_s}+\frac{1}{Y_w}\right)^2\right)^{3/2}\right] \quad (41)$$

From (41), we know that $r'>0$, indicating that $r$ increases with $Y_s$. Since $r<0$, as $Y_s$ (and thus $\alpha$) increases, the impact of active current injection from GFLR on system voltage weakens.

In summary, as $\alpha$ increases, $U_w$ becomes less sensitive to changes in the active current of GFLRs, while the sensitivity of SG's rotor angle dynamics to GFLR current changes increases. At $\alpha\approx 0.5$, $U_w$ is most sensitive to SG's rotor angle changes. Therefore, the study focuses on the range $\alpha\in(0.5,1)$, where interactions between SG and GFLR are more pronounced, making SG rotor angle stability the primary concern.

### B. Proof of the Multi-Swing Stability Boundary being an Unstable Limit Circle

**Theorem 1**(*Poincare Annular Region Theorem*)[32]: If **D** is an annular region that does not contain any equilibrium point, and any trajectory that intersects the boundary of **D** moves in the exterior-to-interior (interior-to-exterior) direction, then there exists at least one stable (unstable) limit cycle. If the inner boundary of **D** shrinks into an unstable (stable) point, the theorem can still be established.

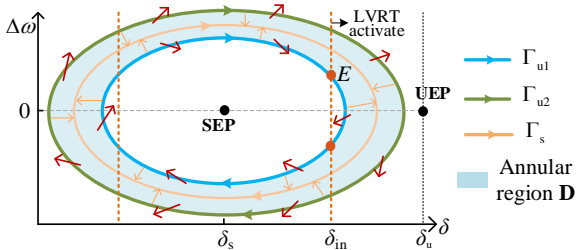

**Fig. 18.** Multi-swing stability boundary of the system

Let $\Gamma_{u1}$ be the true stability boundary of the system. Due to the control dynamics of the GFLR after fault clearance, transient energy may continue to accumulate, narrowing the SG's stability boundary compared to the first-swing boundary through the UEP. As shown in Fig. 18, the maximum rotor angle on $\Gamma_{u1}$ is less than the angle $\delta_u$ corresponding to the UEP. Thus, there exists an annular region **D** containing $\Gamma_{u1}$, and **D** does not contain the system's equilibrium points (SEP and UEP). Any trajectory crossing these boundaries exits **D**, as indicated by the red arrows in Fig. 18. According to Theorem 1, there is at least one unstable limit cycle within **D**. If there is only one unstable limit cycle, $\Gamma_{u1}$ serves as the system's stability boundary. If there are multiple unstable limit cycles ($\Gamma_{u1}$ and $\Gamma_{u2}$), as shown in Fig. 18, a stable limit cycle ($\Gamma_s$) must exist between them. If the system stabilizes on $\Gamma_s$, it will oscillate, which is considered unstable in terms of Lyapunov asymptotic stability and practical engineering. Therefore, $\Gamma_{u1}$ remains the system's stability boundary and is an unstable limit circle.

### C. Test System Parameters

TABLE IV. PARAMETERS OF TEST SYSTEM 1

| | Parameter Name | Parameter Value |
|---|---|---|
| Wind Farms | Total capacity | 1500MVA |
| | Outer loop proportional/integral coefficient ($k_{ip}, k_{io}$) | GSC 1,3,4: 0.2, 4 GSC 2,5: 0.7, 0.14 |
| | SCR for grid connection | GSC 1-5: 2.716; 2.261; 3.928; 2.173; 1.868 |
| | LVRT reactive compensation coefficient ($K_q$) | 2 |
| SGs | Capacity | 2520MVA |
| | Transient d,q-axis reactance | 0.28,0.39 |
| | Inertia time constant ($T_J$) | 7s |
| Grid-connected branch | WF connection branch($1/Y_w$) | 0.18 p.u. |
| | SG connection branch($1/Y_s$) | 0.128p.u. |
| | Branch after PCC($1/Y_g$) | 0.26p.u. |
| Receiving-end grid | Voltage level | 525kV |
| | Base frequency | 50Hz |
| | Base capacity | 1000MVA |

TABLE V. PARAMETERS OF TEST SYSTEM 2

| | Parameter Name | Parameter Value | |
|---|---|---|---|
| GFLRs | Outer loop proportional/integral coefficient ($k_{po}, k_{io}$) | GFLR1: 2,10 | GFLR2: 2,10 |
| | Inner loop proportional/integral coefficient ($k_{pi}, k_{ii}$) | GFLR1: 50, 200 | GFLR2: 100,250 |
| | PLL proportional/integral coefficient ($k_{pl}, k_{il}$) | GFLR1: 10, 100 | GFLR2: 15, 100 |
| | Filter inductor, capacitor, resistor | 8e$^{-3}$ H, 1e$^6$ μF, 5e$^{-3}$ Ω | |
| SG | Capacity, steady state output | 120MVA, 95MW | |
| | Transient reactance | 0.15 p.u. | |
| | Physical damping coefficient ($D$) | 3 p.u./p.u. | |
| Branches | $Y_s, Y_w, Y_g$ | 5, 5, 1.67 p.u. | |
| Grid | Voltage level | 100kV | |
| | Base frequency and capacity | 50Hz, 1000MVA | |

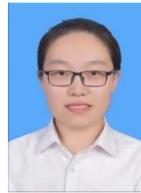
**Bingfang Li** (Graduate Student Member, IEEE), received the B.S. degree from North China Electric Power University, Baoding, China, in 2022, and is currently working toward the Ph.D. degree with Xi'an Jiaotong University. Her main fields of interest include Power system stability analysis and control.

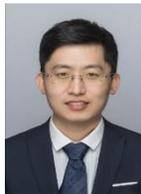
**Songhao Yang** (Senior Member, IEEE) was born in Shandong, China, in 1989. He received the B.S. and Ph.D. degrees in electrical engineering from Xi'an Jiaotong University, Xi'an, China, in 2012 and 2019, respec-tively, and the Ph.D. degree in electrical and electronic engineering from Tokushima University, Tokushima, Japan, in 2019. He is currently an Associate Professor with Xi'an Jiaotong University. His research focuses on power system stability analysis and control.

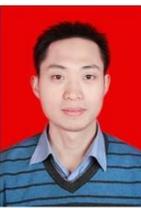
**Guosong Wang**, born in Guizhou Province, China, in 1978, received his bachelor's degree from Wuhan University of Hydraulic and Electric Engineering in 2001. He has long been engaged in power system dispatch and operation, security and stability analysis, and security and stability control.

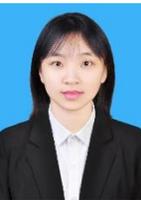
**Yiwen Hu**, received the B.S. degree from North China Electric Power University, Baoding, China, in 2023, and is currently working toward the M.S. degree with Xi'an Jiaotong University. Her main fields of interest include Power system stability analysis and control.

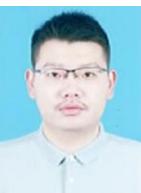
**Xu Zhang** (Graduate Student Member, IEEE), received the B.S. degree from Xi'an Jiaotong University, Xi'an, China, in 2021, and is currently working toward the Ph.D. degree with Xi'an Jiaotong University. Her main fields of interest include Power system voltage stability analysis.

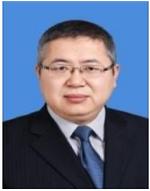

**Zhiguo Hao** (Senior Member, IEEE), was born in Ordos, China, in 1976. He received the B.Sc. and Ph.D. degrees in electrical engineering from Xi'an Jiaotong University, Xi'an, China, in 1998 and 2007, respectively. He is currently a Professor with the Electrical Engineering Department, Xi'an Jiaotong University. His research focuses on power system protection and control.

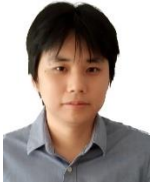

**Dongxu Chang**, (Member, IEEE) was born in He-nan, China, in 1982. He received the B.S. degrees in electrical engineering from HUST, Wuhan, China, in 2006. He is currently working at the Electric Power Research Institute，CSG. His research focuses on power system stability analysis and control.

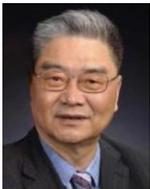

**Baohui Zhang** (Fellow, IEEE) was born in Hebei Province, China, in 1953. He received the M.Eng. and Ph.D. degrees in electrical engineering from Xi'an Jiaotong University, Xi'an, China, in 1982 and 1988, respectively. Since 1992, he has been a Professor with Electrical Engineering Department, Xi'an Jiaotong University. His research interests are system analysis, control, communication, and protection.